\pdfoutput=1
\documentclass[12pt]{article}
\usepackage{graphicx}

\usepackage[a4paper, twoside,
            top=2.2cm,bottom=2.2cm,
            left=2.0cm,right=2.0cm,
            ]{geometry}

\newcommand\pubnumber{}
\newcommand\pubdate{\today}

\def\napoli{The Czech Academy of Sciences, Institute of Physics} 

\def\Title#1{\begin{center} {\Large #1 } \end{center}}
\def\Author#1{\begin{center}{ \sc #1} \end{center}}
\def\Address#1{\begin{center}{ \it #1} \end{center}}

\newcommand\pubblock{\rightline{\begin{tabular}{l} \pubnumber\\
         \pubdate  \end{tabular}}}
\newenvironment{Abstract}{\begin{quotation}  }{\end{quotation}}
\newenvironment{Presented}{\begin{quotation} \begin{center} 
             PRESENTED AT\end{center}\bigskip 
      \begin{center}\begin{large}}{\end{large}\end{center} \end{quotation}}
\def\Acknowledgements{\bigskip  \bigskip \begin{center} \begin{large}
             \bf ACKNOWLEDGEMENTS \end{large}\end{center}}

\def\beq{\begin{equation}}
\def\eeq#1{\label{#1}\end{equation}}
\def\eeqn{\end{equation}}

\def\beqa{\begin{eqnarray}}
\def\eeqa#1{\label{#1}\end{eqnarray}}
\def\eeqan{\end{eqnarray}}

\let\bar=\overbar

\def\Dslash{\not{\hbox{\kern-4pt $D$}}}
\def\dslash{\not{\hbox{\kern-2pt $\del$}}}

\def\msb{{\bar{\ssstyle M \kern -1pt S}}}

\RequirePackage[colorlinks,citecolor=blue,urlcolor=blue,linkcolor=blue]{hyperref}
\usepackage{mathtools}
\usepackage{xifthen}
\usepackage{keycommand}
\usepackage{amssymb,amsmath,latexsym}
\RequirePackage{mathptmx}      

\usepackage[labelfont={small,bf},textfont=small]{caption}
\usepackage[labelfont={small,bf},textfont=small]{subcaption}

\usepackage{cleveref}  
\newcommand{\refp}[1]{(\ref{#1})}

\crefname{equation}{eq.}{eqs.}
\crefname{section}{sect.}{sects.}
\crefname{chapter}{chapter}{chapters}
\crefname{table}{table}{tables}
\crefname{figure}{fig.}{figs.}
\crefname{appsec}{appendix}{appendices}
\crefname{appchap}{appendix}{appendices}

\DeclareMathOperator{\e}{e}

\newcommand{\abs}[1]{\ensuremath{\left| {#1} \right|}}
\newcommand{\ampl}[2][]{\ensuremath{F_{#1}^{\text{#2}}(s,t)}}
\newcommand{\modulus}[2][]{\ensuremath{\abs{\ampl[#1]{#2}}}}
\newcommand{\phase}[1][]{\ensuremath{\zeta^\text{N}_{#1}(s,t)}}

\newcommand{\dcss}[3][]{
\ifthenelse{\equal{#1}{}}
{\ensuremath{      \frac{\text{d}\sigma^{#2}{#3}}{\text{d}t}}}
{\ensuremath{\left.\frac{\text{d}\sigma^{#2}{#3}}{\text{d}t}\right|_{#1}}}
}

\newcommand{\dcs}[2][]{\dcss[#1]{#2}{}}

\newcommand{\weight}[0]{w}
\newkeycommand{\meanb}[n=1,etype=,j=,weight=]{\ensuremath{\langle \ifthenelse{\equal{\commandkey{n}}{1}}{b}{b^{\commandkey{n}}} \rangle^{\text{\commandkey{etype}}\ifthenelse{\equal{\commandkey{weight}}{}}{}{,{\text{\weight}=\commandkey{weight}}}}_{\commandkey{j}} }}
\newkeycommand{\bmax}[j=]{ \ensuremath{b^{\text{max}}\ifcommandkey{j}{_{\commandkey{j}}}{} }}

\newkeycommand{\PROB}[etype=,j=]{ \ensuremath{\ifcommandkey{etype}{P^{\text{\commandkey{etype}}}}{P}_{\ifcommandkey{j}{\commandkey{j}}{}} }}

\newkeycommand{\CS}[etype=,j=]{ \ensuremath{\ifcommandkey{etype}{\sigma^{\text{\commandkey{etype}}}}{\sigma}_{\ifcommandkey{j}{\commandkey{j}}{}} }}

\newcommand{\TITLE}{Eikonal model analysis of elastic proton-proton collisions \\ at 52.8 GeV and 8 TeV}

\newcommand{\pathToFigs}[0]{./}

\begin{document}
\hypersetup{pageanchor=false}
\begin{titlepage}
\pubblock

\vfill
\Title{\TITLE}
\vfill
\Author{Ji\v{r}\'{\i} Proch\'{a}zka\footnote{jiri.prochazka@fzu.cz}, Vojt\v{e}ch Kundr\'{a}t\footnote{kundrat@fzu.cz}, M.~V.~Lokaj\'{\i}\v{c}ek\footnote{lokaj@fzu.cz}}
\Address{\napoli}
\vfill
\begin{Abstract}
Under the influence of standardly used description of Coulomb-hadronic interference proposed by West and Yennie the protons have been interpreted as transparent objects; elastic events have been interpreted as more central than inelastic ones. It is known that using more general eikonal model the measured elastic data may be interpreted also very differently; elastic processes being more peripheral than inelastic ones. The most ample elastic data set measured at ISR energy of 52.8~GeV have been recently reanalyzed with the help of the eikonal model and new results obtained. 
The impact of recently established electromagnetic form factors on determination of quantities specifying hadron interaction determined from the fits of experimental elastic data have been studied. The influence of some other assumptions on proton characteristics derived from elastic hadronic scattering amplitude determined on the basis of experimental data have been analyzed, too. It concerns mainly the assumed $t$-dependence of phase of elastic hadronic amplitude. The results may be then compared to similar analysis of experimental data at much higher LHC energy of 8~TeV recently published by TOTEM experiment.
\end{Abstract}
\vfill
\begin{Presented}
Presented at EDS Blois 2017, Prague, \\ Czech Republic, June 26-30, 2017
\end{Presented}
\vfill
\end{titlepage}
\def\thefootnote{\fnsymbol{footnote}}
\setcounter{footnote}{0}
\hypersetup{pageanchor=true}

\section{\label{sec:introduction}Introduction}
The $t$-dependence of differential cross section for elastic scattering of two charged hadrons (protons) can be defined with the help of complete elastic scattering amplitude $\ampl{C+N}$ at all measured $t$ values as
\begin{equation}
\frac{\text{d} \sigma}{\text{d}t} = \frac{\pi}{s p^2} \modulus{}^2.
\label{eq:difamp_gen}
\end{equation}
Here $s$ is the square of the total collision energy, $t$ is the four momentum transfer squared and $p$ is the value of momentum of one incident proton in the center-of-mass system.  According to Bethe \cite{Bethe1958} the complete amplitude is commonly decomposed into the sum of the Coulomb scattering amplitude $\ampl{C}$ and the hadronic amplitude $\ampl{N}$ bound mutually with the help of relative phase $\alpha\phi(s,t)$:
\begin{equation}
\ampl{C+N} = \ampl{C}\e^{\text{i}\alpha\phi(s,t)}+\ampl{N};
\label{eq:FCNbethe} 
\end{equation}
where $\alpha=1/137.036$ is the fine structure constant. West and Yennie (WY) \cite{WY1968} derived for the $t$-dependence of the relative phase $\alpha\phi(s,t)$ the following formula
\begin{equation}
\alpha\phi(s,t) = \mp \alpha\left[\ln\left (\frac{-t}{s}\right) + \int_{-4p^2}^0 \frac{\text{d}t'}{\abs{t-t'}} \left(1-\frac{F^\text{N}(s,t')}{F^\text{N}(s,t)}\right)\right].
\label{eq:phaseWY}
\end{equation}
The upper (lower) sign corresponds to the scattering of particles with the same (opposite) charges.

For practical reasons the analytical integration over all admissible values of $t'$ has been performed provided the following two assumptions concerning the hadronic amplitude $\ampl{N}$, defined as
\begin{equation} 
\ampl{N} = \text{i}\modulus{N} \e^{-\text{i}\phase} \, ,
\label{eq:modphas}
\end{equation}
at all kinematically allowed $t$ values have been accepted: the modulus $\modulus{N}$ has had purely exponential $t$-dependence and the phase $\phase$ has been constant.  As introduced in \cite{KL2005} some other high energy approximations were added, too.

The relative phase has been then simplified to 
\begin{equation}
\alpha \phi(s,t) = \mp \alpha \left [\ln{\left(\frac{-B(s)t}{2}\right)}+\gamma \right]
\label{eq:phiWY}
\end{equation}
where $\gamma=0.577215$ is Euler constant and $B$ is $t$-independent diffractive slope generally defined as
\begin{equation}
B(s,t) = \frac{\text{d}}{\text{d} t} \left[ \ln \dcs{\text{N}}(s,t)\right]
= \frac{2}{\modulus{N}} \frac{\text{d}}{\text{d}t}\modulus{N} \; .
\label{eq:slope}
\end{equation}
The $t$-independence of $B(t)$ is equivalent to the requirement of purely exponential $t$-dependence of \modulus{N}. 

One may further define quantity $\rho(s,t)$ as ratio of the real to the imaginary part of elastic hadronic amplitude 
\begin{equation}
\rho(s,t) = \frac{\Re \ampl{N}}{\Im \ampl{N}}.
\label{eq:rho}
\end{equation}
It follows from \cref{eq:modphas,eq:rho} that
\begin{equation}
\tan{\phase} = \rho(s,t) \; , 
\label{eq:tanzeta}
\end{equation}
i.e., the assumption concerning $t$-independence of hadronic phase $\phase$ is fully equivalent to assumption of quantity $\rho(s,t)$ being $t$-independent.

The complete elastic scattering amplitude $\ampl{C+N}$ has been then written as
\begin{equation} 
\ampl[\text{WY}]{C+N}  =  \pm \frac{\alpha s}{t}G_1(t)G_2(t) \e^{\text{i}\alpha \phi(s,t)} \\
+ \frac{\CS[etype={tot}](s)}{4\pi}p\sqrt{s}(\rho(s)+\text{i})\e^{B(s)t/2}.
\label{eq:simplifiedWY}
\end{equation}
Here $G_1(t)$ and $G_2(t)$ are the electric dipole form factors being put into formula \refp{eq:simplifiedWY} by hand and the quantity $\CS[etype={tot}](s)$ is the total cross section given by the optical theorem
\begin{equation}
  \CS[etype={tot}](s) = \frac{4 \pi}{p\sqrt{s}} \Im F^{\text{N}}(s,t=0) \, .
  \label{eq:optical_theorem}
\end{equation}

Contrary to the fact that the mentioned theoretical assumptions have not been fulfilled by the analyzed data, formulas \refp{eq:difamp_gen}, \refp{eq:simplifiedWY} and \refp{eq:phiWY} have been commonly used for the analysis of all hitherto pp elastic scattering data in the forward region. In the region of higher values of $|t|$ the influence of Coulomb scattering has been then neglected by definition and the scattering has been described only by phenomenologically constructed hadronic amplitude having different $t$-dependence of \ampl{N} than the one assumed in the WY approach; the whole description has been, therefore, inconsistent. This "standard" elastic hadronic amplitude used to describe data at higher values of $|t|$ had (without any reasoning) typically a dominant imaginary part in a quite broad interval of lower $|t|$ values and was equal to zero in the dip region. However, the existence of minimum (dip) in the differential cross section observed practically in all elastic hadron collisions does not require zero value of its imaginary part; \emph{only the sum of the squares of both the real and imaginary parts should be minimal in this region}. The mentioned requirements have represented much stronger and more limiting conditions that the theory and experiment have required. These limiting assumptions have been included in many contemporary models of elastic hadronic amplitude; they have led to central behaviour of elastic scattering \cite{Kundrat1981}, i.e., having lower value of elastic root-mean-square of impact parameter $\sqrt{\meanb[n=2,etype=el]}$ lesser than the inelastic one $\sqrt{\meanb[n=2,etype=inel]}$ \cite{Kundrat2002}. Central behaviour of elastic collisions has never been satisfactorily explained in literature. 

\section{\label{sec:eikonal}Eikonal model approach}
In order to avoid the given discrepancies and limitations another approach based on the eikonal model has been proposed in \cite{Kundrat1994_unpolarized} and recently revisited in \cite{Prochazka2017}. This framework allowed to derive more general formula for complete elastic scattering amplitude describing the influence of both the Coulomb and hadronic scattering in the whole measured region of momentum transfers in consistent way practically for any hadronic phase $t$-dependence (and modulus of \ampl{N}). The complete amplitude in the eikonal model valid at any $s$ and $t$ up to the terms linear in $\alpha$ may be written as
\begin{equation}
\ampl{C+N} = \pm \frac{\alpha s}{t} G^2_{\text{eff}}(t) + \ampl{N}[1 \mp \text{i} \alpha \bar{G}(s,t)],
\label{eq:kl1}
\end{equation}
where
\begin{equation}
\bar{G}(s,t) = \int\limits_{t_{\text{min}}}^0 \text{d}t'
\left\{ \ln \left( \frac{t'}{t} \right) \frac{\text{d}}{\text{d}t'} \left[{ G^2_{\text{eff}}(t') }\right] \right.\\
- \left. \frac{1}{2\pi} \left[ \frac{F^{\text{N}}(s,t')}{\ampl{N}} - 1 \right]
I(t,t')\right\} 
,
\label{eq:kl2}
\end{equation}
and
\begin{equation}
I(t,t')=\int\limits_0^{2\pi}\text{d}{\Phi''}
\frac{G^2_{\text{eff}}(t'')}{t''};
\label{eq:KLampI}
\end{equation}
here $t'' = t+t'+2\sqrt{tt'}\cos{\Phi''}$. The upper (lower) sign corresponds to the scattering of particles with the same (opposite) charges. $G^2_{\text{eff}}$ is effective form factor squared reflecting the electromagnetic structure of colliding protons and has been introduced in \cite{Block1996} as
\begin{equation}
G_{e\!f\!f}^2 (t) = \frac{1}{1 + \tau} \left[ G_{\text{E}}^2(t) +  \tau \; G_{\text{M}}^2(t) \right] \;,
\;\;\;\;\; \tau = -\frac{t}{4 m^2} \,
\label{eq:Geff}
\end{equation}
where $G_{\text{E}}$ and $G_{\text{M}}$ stand for electric and magnetic form factor; $m$ is the proton mass.

As it has been mentioned in \cite{Prochazka2017,Prochazka2015_bdependence} the distribution of elastic hadron scattering in the impact parameter $b$-space can be analyzed with the help of generalized Fourier-Bessel transform which should be consistent with \emph{finite} allowed region of variable $t$ and at \emph{finite} energies \cite{Islam1968}
\begin{equation}
\begin{split}
h_{\text{el}}(s,b) =& h_{1}(s, b) + h_{2}(s, b)  \\
                   =& \frac{1}{4p\sqrt{s}}\int\limits_{-\infty}^{t_\text{min}} \ampl{N} J_0(b\sqrt{-t})\text{d}t
                   + \frac{1}{4p\sqrt{s}}\int\limits_{t_\text{min}}^0 \ampl{N} J_0(b\sqrt{-t})\text{d}t \; .
\label{eq:fb1}
\end{split}
\end{equation}
In this case the unitarity equation in $b$-space is
\begin{equation}
\Im h_1(s,b)  =  |h_1(s,b)|^2 + g_1(s,b) + K(s,b) \; .
\label{eq:fb2}
\end{equation}
Here $g_1(s,b)$ is real inelastic overlap function which has been introduced in similar way as the complex elastic amplitude in \cref{eq:fb1}.  The complex function $h_{1}(s,b)$, and real functions $g_1(s,b)$ oscillate at finite energies. The oscillations can be removed if a real function \mbox{$c(s,b)=-\Im h_{2}(s,b)$} fulfilling some mathematical conditions is added to both sides of unitarity equation \refp{eq:fb2} \cite{Prochazka2017,Prochazka2015_bdependence}.

\Cref{eq:kl1,eq:kl2,eq:KLampI} have been originally \cite{Kundrat1994_unpolarized} used only with electric form factors. However, there is quite significant difference in $t$-dependence between electric and effective electromagnetic form factors, see detailed discussion in sect.~2 in \cite{Prochazka2017}. A question has been, therefore, raised about the impact of inclusion of magnetic moment of colliding protons on determination of hadronic quantities.

Analysis of experimental elastic pp data \cite{Bystricky1980} with the help of \cref{eq:kl1,eq:kl2,eq:KLampI} with either effective electric or effective electromagnetic proton form factors requires a convenient parameterization of the complex elastic hadronic amplitude, i.e., of its modulus and of its phase. 
The modulus may be parameterized as
\begin{equation}
  \modulus{N} \; = \;\;\; (a_1 + a_2t)\e^{b_1t+b_2t^2+b_3t^3} \; \\
                      +   (c_1 + c_2t)\e^{d_1t+d_2t^2+d_3t^3} \;
\label{eq:ampl_n_mod_param} 
\end{equation}
and can be determined quite uniquely from experimental data. However, the $t$-dependence of the hadronic phase $\zeta(s,t)$ cannot be (if the spins of colliding hadrons are not taken into account) uniquely determined from fits of data. 
The phase may be parameterized as 
\begin{equation} 
  \phase = \arctan{\frac{\rho_0}{1-\abs{\frac{t}{t_{\text{dip}}}}}}
  \label{eq:ampl_n_stdphase}
\end{equation}
where $t_{\text{dip}}$ is the position of the dip. This limited parameterization roughly corresponds to $t$-dependence of phase included in many commonly used hadronic models and it leads to central character of elastic collisions.

In order to explicitly demonstrate alternative peripheral behaviour of elastic collisions in the impact parameter space ($\sqrt{\meanb[n=2,etype=el]} > \sqrt{\meanb[n=2,etype=inel]}$) the phase may be parameterized as
\begin{equation} 
  \phase = \zeta_0 + \zeta_1 \abs{\frac{t}{t_0}}^\kappa \e^{\nu t}, \;\;\; t_0 = 1 \; \; \text{GeV}^2 \;. 
\label{eq:ampl_n_zeta_gen}
\end{equation}
This parameterization allows rather fast increase of $\phase$ with increasing $|t|$ which is inevitable for increasing the value of $\sqrt{\meanb[n=2,etype=el]}$  (for detail see, e.g., \cite{Kundrat1981,Kundrat1994_unpolarized,Kundrat2002}). All parameters specifying the modulus and the phase of elastic hadronic amplitude \ampl{N} may be energy dependent. 

Theoretical analyses of elastic hadronic pp amplitudes \cite{Epstein1969,Eden1971,Block1985} have required the hadronic amplitude to be analytic function in $t$ variable; it means that both the modulus and the phase should be analytic functions, too. While the modulus is obviously analytic the peripheral phase is analytic (see Cauchy-Rieman differential conditions in polar coordinates) only provided the exponent $\kappa$ is positive integer. In all our peripheral fits of pp scattering at energy of 52.8~GeV we have chosen $\kappa = 3$. 

Several fits of the same data at 52.8~GeV (taking into account only statistical errors) under different assumptions have been performed by minimizing the corresponding $\chi^2$ function with the help of program MINUIT~\cite{James1975}. In all the fits estimated errors have been determined with the help of HESSE procedure in MINUIT program. \Cref{tab:pp53gev_ff_eff} shows one central and 3 different peripheral descriptions corresponding to effective electromagnetic form factors. The fitted quantities characterizing the elastic hadronic amplitude are only slightly changed when derived with the help of complete elastic amplitude including the new version of proton effective electromagnetic form factors. Greater change concerns only the values of the $\rho$ quantity in forward direction. 

Choice of $t$-dependence of hadronic phase $\phase$ has, however, fundamental impact on character of collisions in $b$-space. The total root-mean-squared impact parameter values $\sqrt{\meanb[n=2,etype=tot]}$ in \cref{tab:pp53gev_ff_eff} are quite similar for all the 4 alternatives and are equal to approximately 1.02~fm, while the values of the root-mean-squares for elastic and inelastic processes differ significantly. In the central case it holds \mbox{$\sqrt{\meanb[n=2,etype=el]} < \sqrt{\meanb[n=2,etype=tot]}$} while in any of the peripheral alternatives the relation is reversed. 

It may be also interesting to note that Martin's theorem \cite{Martin1997} is fulfilled in the used peripheral models, see \cite{Prochazka2017}. 

\begin{table*}
\centering
\resizebox{\textwidth}{!}{
\begin{tabular}{lccccc}
\hline \hline
Fit                             &              & 1b                        & 2b                        & 3b                        & 4b                        \\
Case                            &              & central                   & peripheral                & peripheral                 & peripheral             \\
\hline                                                
$\rho(t\!=\!0)$                 &              & 0.0766    $\pm$ 0.0017    & 0.0822   $\pm$ 0.0017    & 0.0824     $\pm$ 0.0017   & 0.0825    $\pm$ 0.016     \\
$B(t\!=\!0)$                    & [GeV$^{-2}$] & 13.514    $\pm$ 0.050     & 13.414   $\pm$ 0.045     & 13.42      $\pm$ 0.41     & 13.431    $\pm$ 0.044     \\
$\CS[etype=tot]$                & [mb]         & 42.71     $\pm$ 0.15  & 42.86  $\pm$ 0.10  & 42.860   $\pm$ 0.10     & 42.86  $\pm$ 0.095     \\
$\CS[etype=el]$                 & [mb]         & 7.472                     & 7.541                     & 7.544                     & 7.539                     \\
$\CS[etype=inel]$               & [mb]         & 35.23                     & 35.31                     & 35.32                     & 35.32                     \\
$\CS[etype=el]/\CS[etype=tot]$  &              & 0.1750                    & 0.1761                    & 0.1760                    & 0.1759                    \\
$\text{d}\sigma^{\text{N}}/                                        
\text{d}t(t\!=\!0)$          & [mb.GeV$^{-2}$] & 93.74                     & 94.49                     & 94.49                     & 94.49                     \\
\hline                                                            
$\sqrt{\meanb[n=2,etype=tot]} $ & [fm]         & 1.027                     & 1.022                     & 1.022                     & 1.023                     \\
$\sqrt{\meanb[n=2,etype=el]}  $ & [fm]         & 0.6764                    & 1.676                     & 1.794                     & 1.994                     \\
$\sqrt{\meanb[n=2,etype=inel]}$ & [fm]         & 1.086                     & 0.8170                    & 0.7621                    & 0.6487                    \\
\hline                                                                     
${\chi}^2/$ndf                  &              & 323/205                   & 274/203                   & 275/203                   & 276/203                   \\
\hline \hline
\end{tabular}} 
\begin{minipage}[t]{1.\textwidth}
\caption{\label{tab:pp53gev_ff_eff}
Results of the analysis of pp elastic scattering at energy of 52.8~GeV corresponding to central and three different peripheral distributions in the impact parameter space.
}
\end{minipage}
\end{table*}

The values of quantities $\CS[etype=tot]$, $\rho(t\!=\!0)$ and $B(t\!=\!0)$ in \cref{tab:pp53gev_ff_eff} may be compared to similar values
\begin{equation}
\begin{aligned}
\CS[etype=tot] =& \; (42.38 \pm 0.27) \; \text{mb}, \\
\rho(t\!=\!0)  =& \; (0.078 \pm 0.010),   \\
B(t\!=\!0)     =& \; (13.1  \pm 0.2) \; \text{GeV}^{-2};
\label{eq:wy_parameters_amaldi1978}
\end{aligned}
\end{equation}
determined earlier in \cite{Amaldi1978} (see also \cite{Amaldi1977}) on the basis of the simplified WY formula~\refp{eq:simplifiedWY}.

\begin{figure}
\begin{minipage}[t]{0.50\textwidth}
\includegraphics*[width=\textwidth]{\pathToFigs/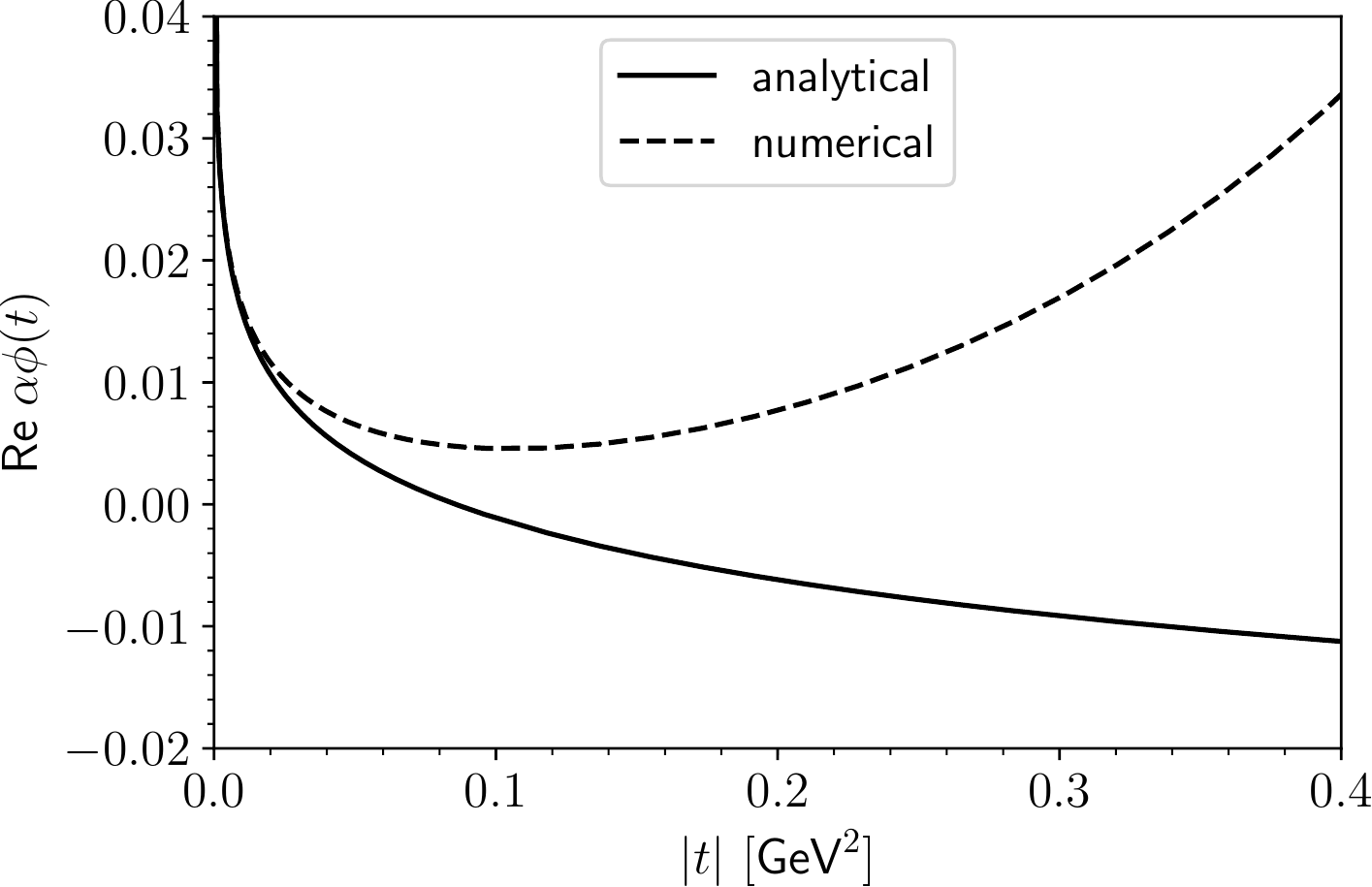}
 \caption{\label{fig:pp53gev_wy_anal_vs_num}Comparison of calculations of the real part of $\alpha\phi(s,t)$ given by \cref{eq:phaseWY} (denoted as "numerical" calculation) and \cref{eq:phiWY} (denoted as "analytical" calculation) 
 under the assumptions of $t$-independent quantities $\rho(t)$ and $B(t)$ whose values have been taken from \refp{eq:wy_parameters_amaldi1978} corresponding to pp scattering at 52.8 GeV. 
}
\end{minipage}
\quad
\begin{minipage}[t]{0.46\textwidth}
\includegraphics*[width=\textwidth]{\pathToFigs/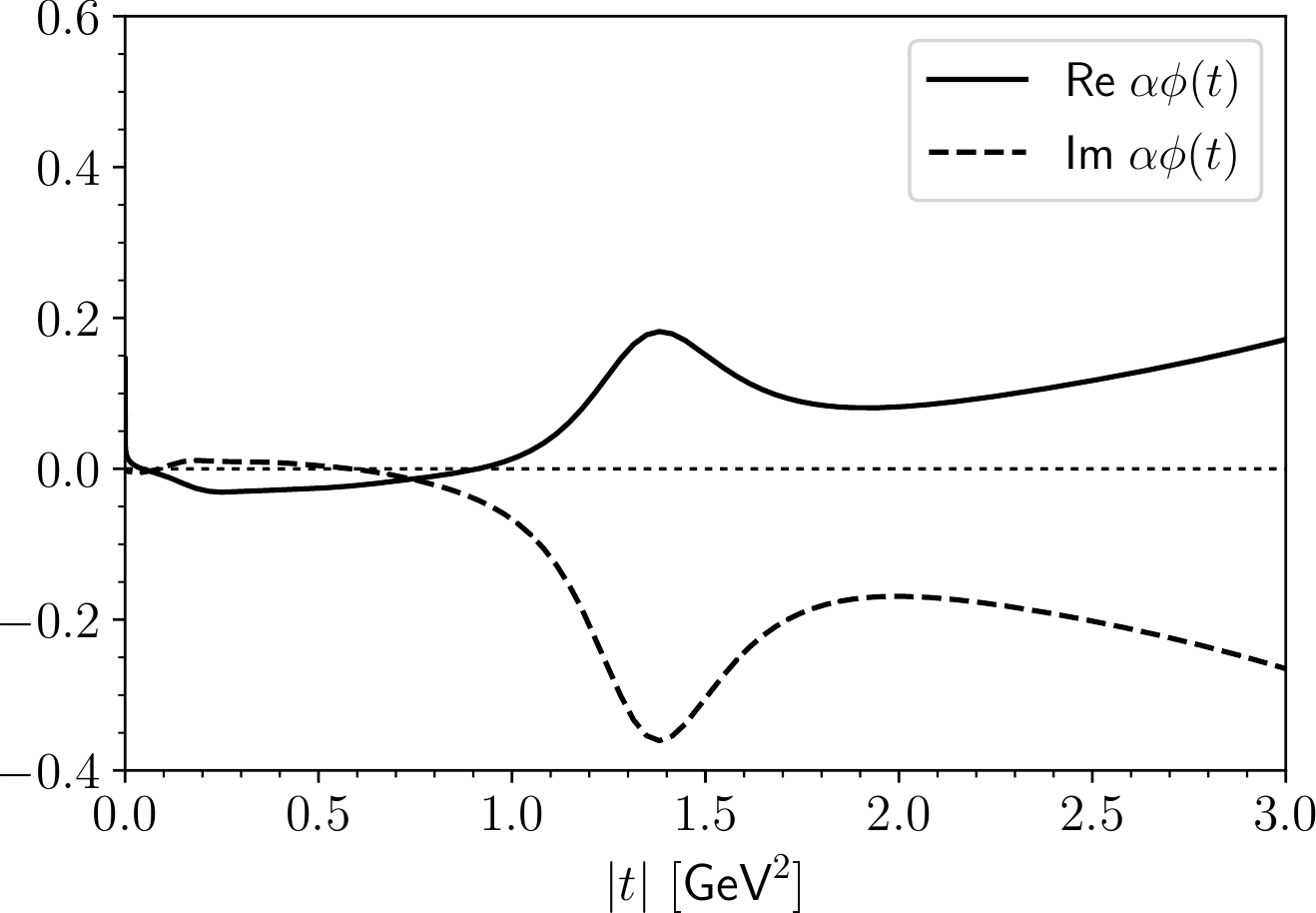}
\caption{\label{fig:pp53gev_wy_phi_phase_re_im}Comparison of the real and imaginary parts of $\alpha\phi(s,t)$ given by \cref{eq:phaseWY} and calculated for elastic pp hadronic amplitude at 52.8~GeV corresponding to 
peripheral Fit~3b. 
}
\end{minipage}
\end{figure}

As it has been mentioned the WY integral formula \refp{eq:phaseWY} has been simplified to \cref{eq:phiWY} under the assumption that the quantities $B(s,t)$ and $\rho(s,t)$ are $t$-independent at all values of $t$. In this case the imaginary part of the relative phase $\alpha \phi(s,t)=0$ by definition. Its real part can be analytically calculated either with the help of \cref{eq:phaseWY} or numerically with the help of \refp{eq:phiWY} under the assumptions of $t$-independent quantities $\rho(t)$ and $B(t)$ whose values have been taken from \refp{eq:wy_parameters_amaldi1978} corresponding to pp scattering at 52.8 GeV\refp{eq:phiWY}; see the comparison in \cref{fig:pp53gev_wy_anal_vs_num}. Both the curves are compatible only for $|t| \lesssim 0.01\; \text{GeV}^2$. 

The WY relative phase integral formula \refp{eq:phaseWY} has to be real (it has been defined as imaginary part of an another function \cite{WY1968}). It has been, however, proved mathematically in \cite{KL2007} that the relative phase defined by \refp{eq:phaseWY} becomes complex if hadronic phase $\phase$ is $t$-dependent. This is illustrated in \cref{fig:pp53gev_wy_phi_phase_re_im} where the $t$-dependence of both the real and imaginary parts of the relative phase $\alpha \phi(s,t)$ corresponding to \ampl{N} with strongly $t$-dependent $\phase$ obtained in peripheral Fit~3b is plotted. The approach of WY a priori strongly limits, without any reasoning, $t$-dependence of hadronic phase and, therefore, it does not allow to perform general studies of $t$-dependence of hadronic phase (and also modulus) on the basis of experimental data.   

\section{\label{sec:conclusion}Conclusion}

The detailed discussion of the results at 52.8~GeV (summarized only very briefly in the preceding) may be found in \cite{Prochazka2017}. Similar analysis of elastic pp scattering data at the LHC energy of 8~TeV has been recently performed with the help of the eikonal model approach in \cite{totem10}. One can find in \cite{Prochazka2017,totem10} the comparison and discussion of different models (corresponding to different assumptions) applied to experimental data. 

We can conclude that the analysis of elastic pp scattering performed with the help of the historically older model of West and Yennie has not been reliable. It has been based on theoretical assumptions which have not been consistent with experimental data and which have a priori constrained $t$-dependence of \ampl{N}, without any reasoning. The description of WY may lead to completely wrong physical conclusions.



The eikonal model has allowed to study $t$-dependence of hadronic amplitude (and, therefore, several properties of hadronic scattering under different assumptions). It may be applied to experimental data at all measured values of $t$. On the basis of the recently performed analyses of experimental data at 52.8~GeV and 8~TeV under different assumptions one may conclude that the choice of form factor (electric vs.~effective electromagnetic) has had small or negligible impact on determination of hadronic quantities. It follows from the performed fits of data that $t$-dependence of hadronic phase has been constrained only weakly by the Coulomb-hadronic interference; some other assumptions need to be added if the hadronic amplitude (including its phase) is to be determined uniquely. It has been explicitly shown for several alternatives that $t$-dependence of hadronic phase has had very strong impact on interpretation of collisions; it may lead to completely different character of collisions in dependence on impact parameter. It should correspond to completely different structures of colliding particles. It is possible to say that there is not any reason against the more realistic interpretation of elastic processes being peripheral and protons regarded as rather compact and not "transparent" objects during elastic collisions.

All the contemporary descriptions of elastic scattering of (charged) hadrons contain open problems and questions. Some of them have been mentioned also during the EDS Blois 2017 conference; see, e.g., \cite{Petrov2016_problems,Petrov2016_sizes}. Several fundamental open problems have been recently summarized by us in \cite{Prochazka2017,Prochazka2015_bdependence}. 
Proper analysis of elastic collisions in dependence on impact parameter may provide important insight concerning shapes and dimensions of collided particles which can be hardly obtained in a different way. 
One should carefully study the assumptions involved in any collision model and test the consequences before making far reaching conclusions concerning structure and properties of collided particles.


\Acknowledgements 
We would like to thank to M.~M.~Islam, A.~K.~Kohara, R.~J.~Luddy, A.~Martin, O.~Nachtmann and V.~A.~Petrov for stimulating discussions concerning various aspects of elastic pp scattering. 

\addcontentsline{toc}{section}{References}

{\footnotesize
\begin{thebibliography}{999}

\bibitem{Bethe1958}
H.~A.~Bethe, Ann.~Phys.~3 (1958) 190.

\bibitem{WY1968} 
G.~B.~West and D.~R.~Yennie, 
Phys.~Rev.~172 (1968) 1413. 

\bibitem{KL2005}
V.~Kundr\'{a}t and M.~Lokaj\'{\i}\v{c}ek,
Phys.~Lett.~B 611 (2005) 102.

\bibitem{Kundrat1981}
V.~Kundr\'{a}t, M.~Lokaj\'{\i}\v{c}ek Jr.~and M.~Lokaj\'{\i}\v{c}ek,
Czech.~J.~Phys.~B 31 (1981) 1334.

\bibitem{Kundrat2002}
V.~Kundr\'{a}t, M.~Lokaj\'{\i}\v{c}ek and D.~Krupa,
Phys.~Lett.~B 544 (2002) 132.

\bibitem{Kundrat1994_unpolarized}
V.~Kundr\'{a}t and M. Lokaj\'{\i}\v{c}ek, Z.~Phys.~C 63 (1994) 619.

\bibitem{Prochazka2017}
J. Proch\'{a}zka, V. Kundr\'{a}t, 
arXiv:1606.09479v3 (2017).

\bibitem{Block1996}
M.~M.~Block, Phys.~Rev.~D 54 (1996) 4337.

\bibitem{Prochazka2015_bdependence}
J.~Proch\'{a}zka, M.~V.~Lokaj\'{i}\v{c}ek and V.~Kundr\'{a}t,
Eur.~Phys.~J.~Plus (2016) 131: 147, see also arXiv:1509.05343.

\bibitem{Islam1968} 
M.~M.~Islam,
Lectures in theoretical Physics, 
ed.~A.~O.~Barut and W.~E.~Brittin, Vol.~10 B
(Gordon and Breach, 1968), 97.


\bibitem{Bystricky1980}
J.~Bystricky et al., in Nucleon-Nucleon and Kaon-Nucleon Scattering,
Landolt-Bornstein, Numerical Data and Functional Relationships in Science and Technology,
New Series, Ed. H.~Schopper, Vol.~9, Springer, Berlin (1980).


\bibitem{Epstein1969}
H.~Epstein, V.~Glaser and A.~Martin, Commun.~Math.~Phys.~12 (1969) 257.

\bibitem{Eden1971} 
R.~J.~Eden, Rev.~Mod.~Phys.~43 (1971) 15.

\bibitem{Block1985}
M.~M.~Block and R.~N.~Cahn, Rev.~Mod.~Phys.~57 (1985) 563.

\bibitem{James1975}
F.~James and M.~Roos, 
Comput.~Phys.~Commun.~10 (1975) 343.

\bibitem{Martin1997}
A.~Martin, Phys.~Lett.~B 404 (1997) 137.

\bibitem{Amaldi1978}
U.~Amaldi et al.,
Nucl.~Phys.~B~145 (1978) 367.

\bibitem{Amaldi1977}
U.~Amaldi et al.,
Phys.~Lett.~B~66 (1977) 390.

\bibitem{KL2007}
V.~Kundr\'{a}t, M.~Lokaj\'{\i}\v{c}ek and I.~Vrko\v{c},
Phys.~Lett.~B 656 (2007) 182.



\bibitem{totem10}
TOTEM Collaboration, Eur.~Phys.~J.~C~76 (2016) 661; see also arXiv:1610.00603. 

\bibitem{Petrov2016_problems}
V.~Petrov,
arXiv:1608.06806v3 (2016).

\bibitem{Petrov2016_sizes}
V.~Petrov,
arXiv:1611.10217 (2016).

\end{thebibliography}}
\end{document}